\documentstyle{article}

\begin{document}
\baselineskip=18pt
\title{\hspace{11cm}{\small\bf IMPNWU-971219}\\
\vspace{2cm}
The nondynamical r-matrix structure of  the elliptic 
Ruijsenaars-Schneider model with N=2}

\author{
 Bo-yu Hou$^{b}$ and Wen-li Yang$^{a,b}$ 
\thanks{e-mail :wlyang@nwu.edu.cn}
\thanks{Fax    :0086-029-8303511}
\\
\bigskip\\
$^{a}$ CCAST ( World Laboratory ), P.O.Box 8730 ,Beijing 100080, China\\
$^{b}$ Institute of Modern Physics, Northwest 
University, Xian 710069, China
\thanks{Mailing address}}

\maketitle

\begin{abstract}
We demonstrate that in a certain gauge  the elliptic Ruijsenaars-Shneider
model with N=2 admits a nondynamical r-matrix structure and the corresponding
classical r-matrix is the same as that of its non-relativistic counterpart
(Calogero-Moser model ) in the same gauge.The relation between our (classical)
Lax operator and the Lax operator given by Ruijsenaars is also obtained.
\vspace{0.7cm}


\end{abstract}
\section{Introduction}
Ruijsenaars-Schneider (RS) model as a relativistic-invariant generalization 
 of the well-known nonrelativistic Calogero-Moser (CM) model[1], has been 
 constructed and shown to be integrable[2]. It describes a completely 
integrable system of N one-dimensional interacting relativistic particles. 
Its importance lies in the fact that it is related to the dynamics of 
solitons in some integrable relativistic field theories[2---5] and the 
corresponding quantum model is realized in terms of commuting difference 
operators--- the Macdonald operator and its elliptic generalization[8,9].
Recent development was shown that it can be obtained by a Hamiltonian 
reducation of the cotangent bundle of some Lie group[6,7].Among all type 
RS model, the elliptic RS model is the most general one and the other type 
such as the rational,hyperbolic and trigonometric case is just the various 
degenerations of the elliptic one.So, the study of the elliptic RS model 
is of great importance in the completely integrable relativistic particles  
system.

The Lax operator of the elliptic RS model was found by Ruijsenaars[2]. On the 
 classical level, the r-matrix structure was constructed in [10] for the 
 degeneration case (the rational, hyperbolic and trigonometric case).The 
 r-matrix structure for the most general case ---elliptic RS model was found  
 in[11].There exists a specific feacture that the fundamental Poisson 
 bracket of the Lax operator is given in terms of a quadratic bracket (or 
 Sklyanin bracket)[13] and the corresponding r-matrix is shown to be of 
 dynamical type.Consequently, besides some difficulties presented by the 
 dynamical aspect of the r-matrix[11,12],  the dynamical Yang-Baxter type 
 relations for RS model is still an open problem[11]. To overcome the above   
 problems may be that whether a ``good" Lax operator for the RS model which 
 has a nondynamical r-matrix structure could be found. In our former work[12], 
 we found such a ``good" Lax operator for the elliptic $A^{(1)}_{N-1}$ 
 CM model. Our main purpose in this paper will be to construct such a ``good"
 Lax operator for the elliptic RS model in the case of N=2
 (i.e the two particle system).In this
 new Lax representation (cf. that of Ruijsenaars'), we find that its  
 r-matrix is of numerical type (nondynamical type) and is the same as that 
 of its nonrelativistic version--- CM model[12].The corresponding r-matrix  
 satisfies the classical Yang-Baxter equation.

\section{Review of the RS model}
The Ruijsenaars-Schneider model is the system of N one-dimensional 
relativistic  particles interacting by the two-body potential.
In terms of the canonical variables $p_i,q_i\ \ (i=1,...N)$ enjoying in
the canonical Poisson bracket
\begin{eqnarray*}
\{p_i,p_j\}=0\ \ ,\ \ \{q_i,q_j\}=0\ \ ,\ \ \{p_i,q_j\}=\delta_{ij}
\end{eqnarray*}
the Hamiltonian of the system is expressed as[2]
\begin{eqnarray}
H=mc^2\sum_{j=1}^{N}coshp_j\prod_{k\neq j}
\left\{\frac{\sigma(q_{jk}+\gamma)\sigma(q_{jk}-\gamma)}{\sigma^2(q_{jk})}
\right\}^{\frac{1}{2}}\ \ \ ,\ \
q_{jk}=q_j-q_k
\end{eqnarray}
\noindent Here, $m$ denotes the particle mass, $c$ denotes the speed of
light, $\gamma $ is the coupling constant and $\sigma(u)$ is some elliptic
function defined in Eq.(9). The Hamiltonian Eq.(1) is known to be
completely integrable[2]. The most effective way to show its integrability
 is to construct the Lax representation for the system (namely,
 to find the Lax operator ,or the classical L-operator).One Lax
 representation
 for the elliptic RS model was first given by Ruijsenaars[2]
 \begin{eqnarray}
 L_{R}(u)^{i}_{j}=\frac{e^{p_j}\sigma(\gamma+u+q_{ji})}{\sigma(\gamma+q_{ji})
 \sigma(u)}\prod_{k\neq j}^{N}
\left\{\frac{\sigma(q_{jk}+\gamma)\sigma(q_{jk}-\gamma)}
{\sigma^2(q_{jk})}\right\}^{\frac{1}{2}} \ \ ,\ \ i,j=1,...,N
\end{eqnarray}
\noindent In this paper, we restrict ourselves to the case $N=2 $. Then ,the
Lax operator given by Ruijsenaars can be written as
\begin{eqnarray}
L_{R}(u)=\left(
\begin{array}{cc}
\frac{e^{p_1}\sigma(\gamma+u)}{\sigma(u)\sigma(\gamma)}
\left\{\frac{\sigma(q_{12}+\gamma)\sigma(q_{12}-\gamma)}
{\sigma^2(q_{12})}\right\}^{\frac{1}{2}}&
\frac{e^{p_2}\sigma(\gamma+u+q_{21})}{\sigma(u)\sigma(q_{21})}
\left\{\frac{\sigma(q_{21}-\gamma)}
{\sigma(q_{21+\gamma})}\right\}^{\frac{1}{2}}\\
\frac{e^{p_1}\sigma(\gamma+u+q_{12})}{\sigma(u)\sigma(q_{12})}
\left\{\frac{\sigma(q_{12}-\gamma)}
{\sigma(q_{12+\gamma})}\right\}^{\frac{1}{2}}&
\frac{e^{p_2}\sigma(\gamma+u)}{\sigma(u)\sigma(\gamma)}
\left\{\frac{\sigma(q_{21}+\gamma)\sigma(q_{21}-\gamma)}
{\sigma^2(q_{21})}\right\}^{\frac{1}{2}}
\end{array}\right)
\end{eqnarray}
\noindent Here, we adpot another Lax operator $\stackrel{\sim}{L}_{R}(u)$,
 which  is the same as that of Nijhoff et al in Ref.[11]
\begin{eqnarray}
\stackrel{\sim}{L}_{R}(u)=\left(
\begin{array}{cc}
\frac{e^{p_1}\sigma(\gamma+u)\sigma(q_{12}+\gamma)}
{\sigma(u)\sigma(\gamma)\sigma(q_{12})}&
\frac{e^{p_2}\sigma(\gamma+u+q_{21})}
{\sigma(u)\sigma(q_{21})}\\
\frac{e^{p_1}\sigma(\gamma+u+q_{12})}
{\sigma(u)\sigma(q_{12})}&
\frac{e^{p_2}\sigma(\gamma+u)\sigma(\gamma+q_{21})}
{\sigma(u)\sigma(\gamma)\sigma(q_{21})}
\end{array}\right)
\end{eqnarray}
\noindent The relation of $\stackrel{\sim}{L}_{R}(u)$ with the standard
Ruijsenaars $L_{R}(u)$ can be obtained from the following canonical
transformation (or Poisson map)
\begin{eqnarray}
q_i\longrightarrow q_i\ \ ,\ \ p_i\longrightarrow p_i+\frac{1}{2}
ln\prod_{k\neq i}\frac{\sigma(q_{ik}+\gamma)}{\sigma(q_{ik}-\gamma)}
\end{eqnarray}
\noindent The fundamental Poisson bracket of the Lax operator
$\stackrel{\sim}{L}_{R}(u)$  can be given in the following quadratic
r-matrix form [11,13]
\begin{eqnarray}
& &\{\stackrel{\sim}{L}_{R}(u)_1,\stackrel{\sim}{L}_{R}(v)_2\}
=\stackrel{\sim}{L}_{R}(u)_1\stackrel{\sim}{L}_{R}(v)_2 r^{-}_{12}(u,v)
-r^{+}_{12}(u,v)\stackrel{\sim}{L}_{R}(u)_1\stackrel{\sim}{L}_{R}(v)_2
\nonumber\\
& &\ \ \ \ +\stackrel{\sim}{L}_{R}(u)_1s^{+}_{12}(u,v)\stackrel{\sim}{L}
_{R}(v)_2
-\stackrel{\sim}{L}_{R}(v)_2s^{-}_{12}(u,v)\stackrel{\sim}{L}_{R}(u)_1
\end{eqnarray}
\noindent where
\begin{eqnarray*}
& & r^{-}_{12}(u,v)=a_{12}(u,v)-s_{12}(u)+s_{21}(v)\ \ ,\ \
r^{+}_{12}(u,v)=a_{12}(u,v)+u^{+}_{12}+u^{-}_{12}\\
& & s^{+}_{12}(u,v)=s_{12}(u)+u^{+}_{12}\ \ \ ,\ \ \
 s^{-}_{12}(u,v)=s_{21}(v)-u^{-}_{12}
\end{eqnarray*}
\noindent and
\begin{eqnarray*}
& & u^{\pm}_{12}=\sum_{i,j}\xi (q_{ji}\pm \gamma)e_{ii}\otimes e_{jj}\ \ \ ,\ \
a_{12}(u,v)=r^{0}_{12}(u,v)+\sum_{i=1}^{2}\xi(u-v)e_{ii}\otimes e_{ii}
+\sum_{i\neq j}\xi(q_{ij})e_{ii}\otimes e_{jj}\\
& & r^{0}_{12}(u,v)=\sum_{i\neq j}\frac{\sigma(q_{ij}+u-v)}{\sigma(q_{ij})
\sigma(u-v)}e_{ij}\otimes e_{ji}\ \ \ ,\ \ \
s_{12}(u)=\sum_{i,j}\left( \stackrel{\sim}{L}_{R}(u)\partial_{\gamma}
\stackrel{\sim}{L}_{R}(u)\right)^{i}_{j}e_{ij}\otimes e_{jj}
\end{eqnarray*}
\noindent $\xi(x)$ is some elliptic function defined in Eq.(9).
The matrix element of $e_{ij}$ is equal to $(e_{ij})^{l}_{k}
=\delta_{il}
\delta_{jk}$.It can be checked that the following symmetric condition
hold for the r-matrices $r^{\pm}_{12}(u,v)$ and $s^{\pm}_{12}(u,v)$
\begin{eqnarray}
& & r^{\pm}_{21}(v,u)=-r^{\pm}_{12}(u,v)\ \ \ ,\ \ \
s^{+}_{21}(v,u)=s^{-}_{12}(u,v)\\
& & r^{+}_{12}(u,v)-s^{+}_{12}(u,v)=r^{-}_{12}(u,v)-s^{-}_{12}(u,v)
\end{eqnarray}
It can be see that the classical r-matrices $r^{\pm}_{12}(u,v)$,
$s^{\pm}_{12}(u,v)$ are of  dynamical ones (i.e the matrix element of theirs
do depend upon the dynamical variables $q_i$).The quadratic Poisson bracket
Eq.(6) and the the symmetric conditions Eq.(7)---Eq.(8) lead to the evolution
integrals $tr(\stackrel{\sim}{L}_{R}(u))^{n}$ of the motion.

Due to the r-matrices depending on the dynamical variables, the Poisson
bracket of $\stackrel{\sim}{L}_{R}(u)$ is no longer closed . The
complexity of the r-matrices Eq.(6) results in that it is still an open
problem to check the generalized Yang-Baxter relations for the RS model.
The same situation also  occur for the standard Lax operator $L_{R}(u)$ [11].

\section{ The new Lax representation for RS model and its nondynamical
r-matrix structure}
The dynamical r-matrix structure for the Lax operator of $L_{R}(u)$ and its
Poisson-equivalence $\stackrel{\sim}{L}_{R}(u)$ for the elliptic RS model
lead to some difficults[11] in the investigation of the RS model :
the Poisson algebra of the Lax operator is no longer closed and the
generalized Yang-Baxter relations is still an open problem.However,
the choice of a gauge for the Lax operator is quite important in that it
influnces to a great extent the complexity of the associated r-matrix
structure[12].This motivate us to find a new Lax representation of the
RS model.Following the success in finding a ``good" Lax operator of the
elliptic CM model, we construct a new Lax operator ( we call it as
a ``good" Lax operator in the sense that it has a nondynamical r-matrix
structure).When taking the nonrelativistic limit of this Lax operator, we
 can obtain the ``good" Lax operator for the corresponding CM model[12].

First,let us define some elliptic functions
\begin{eqnarray}
& &\theta^{(j)}(u)=
\theta\left[\begin{array}{c}\frac{1}{2}-\frac{j}{2}\\ 
\frac{1}{2}\end{array}\right](u,2\tau)\nonumber\\
& &\sigma(u)=\theta\left[\begin{array}{c}\frac{1}{2}\\ 
\frac{1}{2}\end{array}\right](u,\tau)\\
& &\theta\left[\begin{array}{c}a\\ b\end{array}\right](u,\tau)
=\sum_{m=-\infty}^{\infty}exp\{\sqrt{-1}\pi[(m+a)^{2}\tau + 2(m+a)(z+b)]\}
\nonumber\\
& &\theta'^{(j)}(u)=\partial_{u}\{\theta^{(j)}(u)\}\ \ ,\ \
\xi(u)=\partial_{u}\{ln\sigma(u)\}
\nonumber
\end{eqnarray}
\noindent where $\tau$ is a complex number with $Im(\tau)>0$ .We find that 
there exist another Lax representation for the RS model and denote it by  
$L(u)$
\begin{eqnarray}
L(u)^{i}_{j}=\sum_{k=1}^{2}\frac{1}{\sigma(\gamma)}A(u+2\gamma;q)^i_k
A^{-1}(u;q)^k_je^{p_k}\ \ \ ,\ \ i,j=1,2
\end{eqnarray}
\noindent where the matrix $A(u;q)$ is
\begin{eqnarray*}
A(u;q)=
\left( \begin{array}{cc}\theta^{(1)}(u+q_{12}+\frac{1}{2})&
\theta^{(1)}(u+q_{21}+\frac{1}{2})\\
\theta^{(2)}(u+q_{12}+\frac{1}{2})&\theta^{(2)}(u+q_{21}+\frac{1}{2})
\end{array}
\right)
\end{eqnarray*}
The new Lax operator given by us in Eq.(10) can be obtained from the
$\stackrel{\sim}{L}_{R}(u)$ through a dynamical similarity transformation
as follows
\begin{eqnarray}
L(u)=g(u)\stackrel{\sim}{L}_{R}(u)g^{-1}(u)
\end{eqnarray}
\noindent where
\begin{eqnarray*}
g(u)=
\left( \begin{array}{cc}\theta^{(1)}(u+q_{12}+\frac{1}{2})&
-\theta^{(1)}(u+q_{21}+\frac{1}{2})\\
\theta^{(2)}(u+q_{12}+\frac{1}{2})&-\theta^{(2)}(u+q_{21}+\frac{1}{2})
\end{array}
\right)
\end{eqnarray*}
\noindent Due to the transformation Eq.(11) being dependent up the dynamical
variables $q_i$, the corresponding classical r-matrix structure would be
changed drastically. Through the direct calculation, we find that the
fundamental Poisson of $L(u)$ can be written in the standard quadratic
Poisson-Lie bracket  with a purely numerical r-matrix
\begin{eqnarray}
\{L_{1}(u),L_{2}(v)\}=[r_{12}(u-v),L_{1}(u)L_{2}(v)]
\end{eqnarray}
\noindent and the numerical r-matrix $r(u)$  is the same as that of Lax
operator of the nonrelativistic counterpart---the elliptic CM model given by
us[12]
\begin{eqnarray}
r(u)=\left(\begin{array}{llll}a(u)&&&d(u)\\&b(u)&c(u)&\\
&c(u)&b(u)&\\d(u)&&&a(u)\end{array}\right)
\end{eqnarray}
\noindent and 
\begin{eqnarray*}
& & a(u)=\frac{\theta '^{(0)}(u)}{\theta^{(0)}(u)} -\frac{\sigma '(u)}
{\sigma (u)}\ \ \ ,\ \ \ 
b(u)=\frac{\theta '^{(1)}(u)}{\theta^{(1)}(u)}-\frac{\sigma '(u)}
{\sigma (u)}\\
& & c(u)=\frac{\theta '^{(0)}(0)\theta^{(1)}(u)}
{\theta^{(0)}(u)\theta^{(1)}(0)}     \ \ \ ,\ \ \ 
d(u)=\frac{\theta '^{(0)}(0)\theta^{(0)}(u)}
{\theta^{(1)}(u)\theta^{(1)}(0)}    
\end{eqnarray*}
\noindent where $a(u),b(u),c(u),d(u)$ are all independent upon dynamical 
variable. The numerical r-matrix $r(u)$ defined in Eq.(13) satisfies the 
classical Yang-Baxter equation
\begin{eqnarray}
[r_{12}(u-v),r_{13}(u-\eta)]+[r_{12}(u-v),r_{23}(v-\eta)]
+[r_{13}(u-\eta),r_{23}(v-\eta)]=0
\end{eqnarray}
\noindent and enjoys in the antisymmetric properties 
\begin{eqnarray}
r_{12}(u)=-r_{21}(-u)
\end{eqnarray}
The nondynamical r-matrix $r(u)$ is equivalent to that of Sklyanin in [14]
up to some scalar factor independing on the dynamical variable.

The standard quadratic Poisson-Lie bracket Eq.(12) of the Lax operator $L(u)$
and the numerical r-matrix $r(u)$ enjoying in the classical Yang-Baxter 
equation Eq.(14) and antisymmetry Eq.(15), make it possiple to construct 
the quantum version of Eq.(12)
\begin{eqnarray}
R_{12}(u-v)T_{1}(u)T_{2}(v)=T_{2}(v)T_{1}(u)R_{12}(u-v)
\end{eqnarray}
\noindent where $R_{12}(u-v)$ is the eight-vertex Baxter's R-matrix and 
satisfies the quantum Yang-Baxter equation
\begin{eqnarray}
R_{12}(u-v)R_{13}(u-\eta)R_{23}(v-\eta)=
R_{23}(v-\eta)R_{13}(u-\eta)R_{12}(u-v)
\end{eqnarray}
The quantum version of Lax operator $L(u)$ given by us is the quantum
L-operator $T(u)$ which satisfy the quantum Sklyanin algebra[14]. Moreover,
the classical numerical r-matrix $r(u)$ is the semi-classical limit of the 
quantum Baxter's R-matrix $R(u)$
\begin{eqnarray}
R(u)=1+wr(u)+o(w^{2})\ \ \ ,\ \ {\rm when\ \ the\ \ crossing\ \ parameter\ \ 
} w\longrightarrow 0
\end{eqnarray}
\noindent the r-matrix $r(u)$ given by us could also be obtained from that of
Nijhoff et al through some kind classical  twisting procedure of r-matrix[12].

\section*{Discussions}
In this paper, we only consider the special case of $N=2$ for the RS model. 
However, the results can be generalized to the generic case of $2\leq N$ .
We will present the further results in the further paper.Moreover the same
nondynamical r-matrix structure could be constructed for the rational ,
hyperbolic and trigonometric RS model.

\end{document}